\documentclass[prb,twocolumn,aps,showpacs,fixfloats]{revtex4}
\usepackage{graphicx}
\usepackage{bm}
\usepackage{amsmath,amssymb}
\usepackage{subfigure}
\usepackage{float}
\usepackage{latexsym}
\usepackage{color}
\usepackage{enumerate}
\usepackage{pdfpages}
\usepackage{tikz}
\usepackage{hyperref}
\usepackage{graphicx}
\usepackage{bm}
\usepackage{amssymb} 
\usepackage{amsmath}
\usepackage{subfigure}
\usepackage{relsize}

\begin{document}
\newcommand{\s}{\scriptscriptstyle}
\newcommand{\uu}{\uparrow \uparrow}
\newcommand{\ud}{\uparrow \downarrow}
\newcommand{\du}{\downarrow \uparrow}
\newcommand{\dd}{\downarrow \downarrow}
\newcommand{\ket}[1] { \left|{#1}\right> }
\newcommand{\bra}[1] { \left<{#1}\right| }
\newcommand{\bracket}[2] {\left< \left. {#1} \right| {#2} \right>}
\newcommand{\vc}[1] {\ensuremath {\bm {#1}}}
\newcommand{\tr}{\text{Tr}}
\newcommand{\Trans}{\ensuremath \Upsilon}
\newcommand{\Refl}{\ensuremath \mathcal{R}}

\title{Shape of the zeroth Landau level in graphene with non-diagonal disorder}

\author{Rajesh K. Malla   and M. E. Raikh}

\affiliation{ Department of Physics and
Astronomy, University of Utah, Salt Lake City, UT 84112}

\begin{abstract}
Non-diagonal (bond) disorder in graphene broadens Landau levels (LLs) in the same way as random potential. The exception is the zeroth LL, $n=0$, which is robust to the bond disorder, since it does not mix different $n=0$
states within a given valley. The mechanism of broadening of the $n=0$ LL is the inter-valley scattering. Several
numerical simulations of graphene with bond disorder had established that $n=0$ LL is not only anomalously narrow
but also that its shape is very peculiar with three maxima, one at zero energy, $E=0$, and two others at finite
energies $\pm E$. We study theoretically the structure of the states in $n=0$ LL in the
presence of bond disorder. Adopting the assumption that the
bond disorder is strongly anisotropic, namely, that one type of bonds is perturbed
much stronger than  other two,
allowed us to get an analytic expression for the density of states
which agrees with numerical simulations remarkably well.  On the qualitative level,
our key finding is that delocalization of $E=0$ state has a dramatic
back effect on the density of states near $E=0$. The origin of
this unusual behavior is the strong correlation of eigenstates in
different valleys.

\end{abstract}

\maketitle

\section{Introduction}

Broadening of the Landau levels (LLs) in two-dimensional (2D) electron gas by a random potential
was studied more than a quarter century ago in various limits,
namely,
strong and weak magnetic field, and also short-range and
long-range
disorder.\cite{Ando1974,Baskin1978,Wegner1983,Brezin1984,Larkin1981,Affleck1983,Affleck1984,Apel1987,Chalker1985,Chalker1986,Benedict1987,Shahbazyan1993}

With regard to LLs in graphene,\cite{Ando2002,ReviewOnLLsInGraphene}
the theories of the LL broadening  of Refs.
\onlinecite{Ando1974,Baskin1978,Wegner1983,Brezin1984,Larkin1981,Affleck1983,Affleck1984,Apel1987,Chalker1985,Chalker1986,Benedict1987,Shahbazyan1993}
apply. Recent experimental and theoretical studies
of the LLs in graphene in the presence of disorder
are reported
in Refs. \onlinecite{StormerKim,Geim,JapaneseExperimental,Gornyi2008,DisorderLL}.

There is, however, a situation when the underlying mechanism of the LL broadening in
graphene is distinctively different from that in the 2D gas. The tight-binding Hamiltonian
of the disordered graphene in magnetic field has the form
\begin{equation}
\label{Graphene}
{\hat H}=\sum_{i} V_i c_i^{\dagger} c_i+\sum_{\langle i,j \rangle} \left(t_{i,j}e^{i \theta_{i,j}} c_i^{\dagger}c_j + \text{H.c.} \right),
\end{equation}
where the $\langle i,j \rangle$ correspond to neighboring sites and the sum runs over all the sites. The Peierls phase, $\theta_{i,j}$, is defined in such a way
that the sum of
the phases around a unit cell is equal to the magnetic flux
(in the units of flux quantum) threading the cell.
It follows from Eq. (\ref{Graphene}) that the disorder can be of two
types: randomness in on-site energies, $V_i$, describe the potential disorder,
while the randomness in the hopping integrals $t_{i,j}$ describe the bond disorder
specific for graphene.
To describe the LL broadening due to $V_i$, one can use the
continuous version of the bare Hamiltonian  Eq. (\ref{Graphene})
 \begin{equation}
\label{Continuous}
{\hat H}_0= V(\bm{r})\cdot I +v_{\s F} \bm{\pi}\cdot\bm{\sigma},
\end{equation}
where $\bm{\sigma}$ is the 2D vector whose projections are the  Pauli matrices, and $v_{\s F}$ is the Fermi velocity. The effective momentum operator  is given by $\bm{\pi}=\bm{p}-e \bm{A}/c$, with $\bm{p}$ being the electron momentum, and $\bm{A}=(0,B x)$ is the vector potential with $B$ standing for the uniform magnetic field.

To make a connection to Refs.
\onlinecite{Ando1974,Baskin1978,Wegner1983,Brezin1984,Larkin1981,Affleck1983,Affleck1984,Apel1987,Chalker1985,Chalker1986,Benedict1987,Shahbazyan1993},
the Fourier component, $V_{\bm q}$, of the random potential, $V({\bm r})$,
is expressed through the random energies, $V_i$, used in numerical simulations, as
$\sum_{i} V_i\exp\left(i{\bm q}\cdot {\bm r}_i\right)$.

Unlike the potential disorder, the bond disorder
corresponds to the randomness in $v_{\s F}$ and $B$ and, thus, is called non-diagonal disorder.
Indeed, $v_{\s F}$ is related to the average $\langle t_{i,j}\rangle =t$ as
$v_{\s F}=\frac{{\sqrt 3}ta}{2\hbar}$, where $a$
is the lattice constant.
In this way, the fluctuations of $t_{i,j}$ translate into
the position-dependent $v_{\s F}$. Similarly,
the fluctuations of $\theta_{i,j}$
translate into the position-dependent magnetic field.

The broadening of LLs in graphene due to non-diagonal disorder
has been studied numerically in Refs. \onlinecite{Chakravarty2007,Ando2007,Markos,Sweitzer,Japanese,Pereira,Lewenkopf,Roche2016}.
The results of simulations in all the above papers are consistent with each other.
The most prominent feature of these
results is that the broadening of $n\neq 0$
levels is much stronger than the broadening of the $n=0$ level.
This feature can be easily understood from the continuous Hamiltonian Eq. (\ref{Continuous}).
Indeed, the eigenfunctions of ${\hat H}_0$ are the spinors
\begin{equation}
\label{eigenvector}
\Psi_{n,k}(x,y)=\frac{C_n}{\sqrt{L}}\exp{(iky)}\begin{pmatrix}
sgn(n)(-i)\phi_{|n|-1,k} \\ \phi_{n,k}
\end{pmatrix},
\end{equation}
where $L$ is the normalization length. The constant $C_n$ is equal to $1/2$  for $n\neq0$,
and $C_0=1$.
The functions $\phi_{n,k}(x)$ are the eigenfunctions of the harmonic oscillator
\begin{equation}
\label{LL}
\phi_{n,k}(x)=(2^n n! \sqrt{\pi} \ell_{\s B})^{-1/2} e^{- \frac{\left(x-k \ell_{\s B}^2\right)^2}{2 \ell_{\s B}^2}} H_n\left[\left(x-k \ell_{\s B}^2\right)/\ell_{\s B}\right].
\end{equation}
where $H_n(x)$ is the Hermite polynomial and $\ell_{\s B}=\left(\frac{\hbar c}{eB} \right)^{1/2}$ is the magnetic length.
The difference between $n=0$ LL and other LLs is that the matrix element of non-diagonal disorder between the states
$k$  and $k'$ is zero for $n=0$. This is because one of the two components of the
spinor Eq. (\ref{eigenvector}) is zero. In fact, this matrix element remains zero even when the admixtures
of higher LLs to $n=0$ LL are taken into account, which is the manifestation of the Atiyah-Singer theorem\cite{Atiyah}.
Therefore, the broadening of $n=0$ LL is
absent in the continuous limit. Finite broadening requires virtual transitions with large momentum transfer.
This explains why the width of $n=0$ LL in the
simulations of Refs. \onlinecite{Japanese}, \onlinecite{Lewenkopf} dropped off dramatically
with increasing the correlation radius of $t_{i,j}$.

In all the simulations\cite{Chakravarty2007,Ando2007,Markos,Sweitzer,Japanese,Pereira,Lewenkopf,Roche2016}
the shape of the broadened $n=0$ LL was very nontrivial. It differed from conventional Gaussian
in two respects: (i) it exhibited a shallow {\em minimum} at the center and (ii) it possessed a very narrow peak {\em on top of this minimum}. Neither analytical description nor even theoretical interpretation of this peculiar shape are available. The goal of the present paper is to provide such an interpretation.
In addition to the broadening, the authors of Refs. \onlinecite{Chakravarty2007,Ando2007,Markos,Sweitzer,Japanese,Pereira,Lewenkopf,Roche2016}
studied the localization properties of the $n=0$ eigenstates with non-diagonal disorder.
It was established that there are {\em two} split delocalized states away from the center.
In Ref.~\onlinecite{Markos} it was also found that the third delocalized state with very
 unusual energy dependence of the localization length is present
at the center of LL. Below we will also attempt to interpret this observation.
\begin{figure}
\includegraphics[scale=0.3]{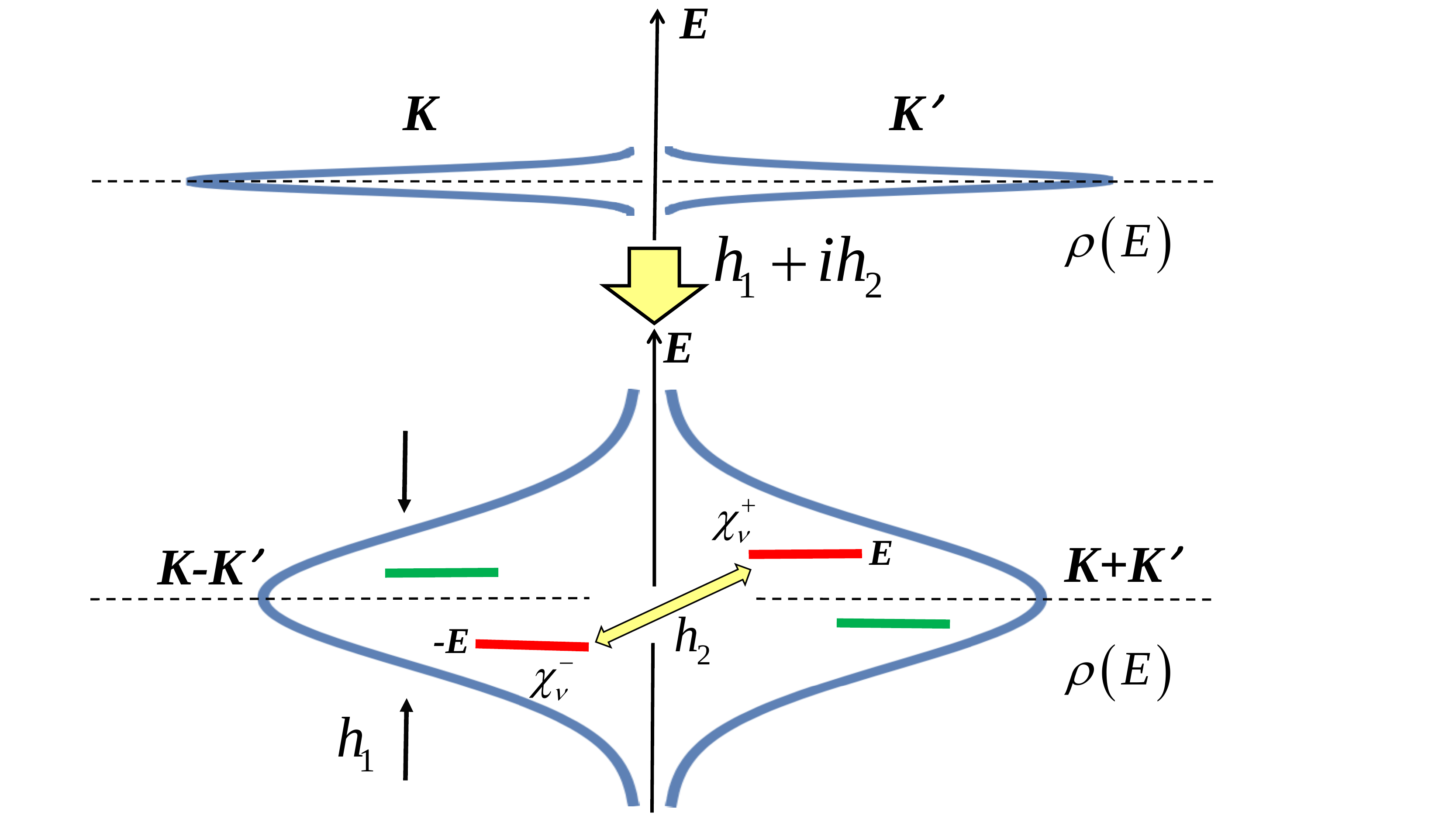}
\caption{(Color online) In order to describe the effect of $K\rightarrow K'$
scattering, described by the nondiagonal term $h_1+ih_2$, on the density of states, we switch to the basis, $K\pm K'$. In the absence of $h_2$, the field $h_1(x,y)$ broadens the  zeroth level, leading to a``gaussian"-type shape of the density of states. Note that, the wavefunction, $\chi_{\nu}^+$, of the state $E$, in the potential $h_1(x,y)$  is the same as the wavefunction,  $\chi_{\nu}^-$, of the state $-E$, in the potential $-h_1(x,y)$, see Eqs. (\ref{System31},~\ref{System32}). A smooth field, $h_2(x,y)$, couples $\chi_{\nu}^+$  to $\chi_{\nu}^-$ (shown with red), but does not couple them to any other states (shown with green). As a result of this coupling, the levels $E$ and $-E$ are repelled away from the center, $E=0$.}
\label{fig1}
\end{figure}

\section{Non-diagonal Disorder}
The way to incorporate the bond disorder into the description of the electron states
was proposed by T. Ando in Ref. \onlinecite{Ando2007}.
One has to add to the bare $4\times 4$ Hamiltonian of graphene
\begin{equation}
\label{Hamiltonian}
{\hat H}_0=v_{\s F}\begin{pmatrix}
0 && \pi_x-\pi_y && 0 && 0 \\ \pi_x+\pi_y && 0 && 0 && 0 \\ 0 && 0 && 0 && \pi_x+\pi_y \\ 0 && 0 && \pi_x-\pi_y && 0
\end{pmatrix}
\end{equation}
a perturbation
\begin{equation}
\label{Andomatrix}
{\hat U}_i(\bm{r})=\begin{pmatrix}
0 && z_A^*z_B && 0 && z_A^*z'_B \\ z_B^*z_A && 0 && z_B^*z'_A && 0 \\
0 && z_A^{'*}z_B &&0 && z_A^{'*}z'_B \\ z_B^{'*}z_A &&0 &&z_B^{'*}z'_A &&0
\end{pmatrix}~u(\bm{r}-\bm{r_i}),
\end{equation}
where $u(\bm{r}-\bm{r_i})$ encodes the change of the hopping parameter
upon the alternation of the bond, $i$. Then the matrix, describing the bond
disorder, takes the form $\sum_i {\hat U}_i$.

Note that\cite{Ando2002} the Hamiltonian Eq. (\ref{Hamiltonian}) represents the continuous limit
of the microscopic Hamiltonian Eq. (\ref{Graphene}),
and its matrix form
captures only the low-energy states close to the points $K$ and $K'$
in the momentum space. The general form of the four-component eigenvectors of ${\hat H}_0$ is $(\psi_A^K,\psi_B^K,\psi_A^{K'},\psi_{B}^{K'})$. In the absence of disorder, the $n=0$ eigenvector has only one nonzero component corresponding to $B$-sites  in the valley $K$ and to $A$-sites in the valley $K'$. The randomness in the hopping parameter couples $B$-sites in the valley $K$
to the $A$-sites in the valley~$K'$.

The fact that there are three types of bonds in graphene is captured in the perturbation ${\hat U}_0$ by non-diagonal matrix element $z_B^*z_A$, where, with proper choice of axes, $(z_B^*z'_A)$ takes three values,
namely, $1$, $\exp(2\pi i/3)$, and $\exp(-2\pi i/3)$, depending on the position of the bond.\cite{Ando2007}

Upon introducing the random function
\begin{multline}
\label{h}
h(\bm{r})=\hspace{-3mm}=\hspace{-1mm}\sum_{bonds~i} \hspace{-3mm}c_i u(\bm{r}-\bm{r_i}) +e^{\frac{2\pi i}{3}}\hspace{-1mm}\sum_{bonds~j} \hspace{-2mm} c_j  u(\bm{r}-\bm{r_j})\\
+e^{-\frac{2\pi i}{3}}\hspace{-1mm}\sum_{bonds~l} \hspace{-3mm}c_l  u(\bm{r}-\bm{r_l}),
\end{multline}
where the coefficients $c_i$, $c_j$, and $c_l$ take the values $0$ or $1$ depending on whether
or not the corresponding bond is perturbed, we rewrite a system of equations for the components of the spinor in the form
\begin{eqnarray}
\label{System}
E\psi_A^K&=&v_{\s F}(\pi_x -i\pi_y)\psi_B^K + h(\bm{r}) \psi_B^{K'}\nonumber\\
E\psi_B^K&=&v_{\s F}(\pi_x +i\pi_y)\psi_A^K + h(\bm{r})\psi_A^{K'}\nonumber\\
E\psi_A^{K'}&=&v_{\s F}(\pi_x +i\pi_y)\psi_B^{K'} + h^{*}(\bm{r})\psi_B^{K}\nonumber\\
E\psi_B^{K'}&=&v_{\s F}(\pi_x -i\pi_y)\psi_A^{K'} + h^{*}(\bm{r})\psi_A^{K}.
\end{eqnarray}
In this system we have kept only the terms responsible for the inter-valley scattering. In the absence of this scattering, the amplitudes $\psi_B^K$ and $\psi_A^{K'}$ correspond to the  $\phi_{0,k}(x)$, i.e. to zeroth LL, while the amplitudes $\psi_A^K$ and $\psi_B^{K'}$ are zero. When the disorder strength is much smaller than the distance between the LLs, the system Eq. (\ref{System}) simplifies to the  $2\times 2$ system
\begin{eqnarray}
\label{System1}
E\psi_B^K&=& h(\bm{r})\psi_A^{K'}\nonumber\\
E\psi_A^{K'}&=& h^{*}(\bm{r})\psi_B^{K}.
\end{eqnarray}
We can write the amplitude $\psi_A^{K'}$ and $\psi_B^K$ as a linear combination of $\phi_{0,k}$,
\begin{eqnarray}
\label{Fourier}
\psi_{B}^{K}=\sum_{\kappa} e^{i\kappa y} \phi_{0,\kappa}(x) B_{\kappa}^K, \nonumber \\
\psi_{A}^{K'}=\sum_{q} e^{i q y} \phi_{0,q}(x) A_{q}^{K'}.
\end{eqnarray}
Substituting Eq. (\ref{Fourier}) into Eq. (\ref{System1}), we get
\begin{eqnarray}
\label{System2}
E B_{\kappa}^K=\sum_q h_{\kappa,q} A_q^{K'},\nonumber \\
E A_{\kappa}^{K'}=\sum_q h_{\kappa,q}^* B_q^{K},
\end{eqnarray}
where $h_{\kappa,q}$ is matrix element of $h({\bm r})$
between the eigenfunctions, $e^{i\kappa y}\phi_{0,\kappa}(x)$ and $e^{iqy}\phi_{0,q}(x)$,
of $n=0$ LL.

\section{Density of states}
\subsection{Perturbative approach}
From Eq. (\ref{Fourier}) it becomes apparent that the problem of
the broadening of $n=0$ LL by the bond disorder reduces
to the model introduced by S. Hikami, M. Shirai, and F. Wegner,
in Ref. \onlinecite{WegnerHikami}.
The Hamiltonian of Ref. \onlinecite{WegnerHikami}
\begin{equation}
\label{HSW}
{\hat H}_{\s HSW}=\begin{pmatrix}
\frac{1}{2m}\left({\bf p}-\frac{e}{c}{\bm A}\right)^2  &&  h_1(x,y)+ih_2(x,y)\\
h_1(x,y)-ih_2(x,y)  &&  \frac{1}{2m}\left({\bf p}-\frac{e}{c}{\bm A}\right)^2
\end{pmatrix},
\end{equation}
pertains to parabolic spectrum with effective mass, $m$.
The random fields, $h_1(x,y)$ and $h_2(x,y)$, are
assumed uncorrelated. When the states are restricted
to zeroth LL, the eigenvectors of the Hamiltonian
${\hat H}_{\s HSW}$ are the spinors with components
\begin{equation}
\label{alphabeta}
\alpha(x,y)=\hspace{-1mm}\sum_{k}\hspace{-1mm} A_k e^{iky}\phi_{0,k}(x,y),~
\beta(x,y)=\hspace{-1mm}\sum_{k}\hspace{-1mm} B_k e^{iky}\phi_{0,k}(x,y),
\end{equation}
where $A_k$ and $B_k$ satisfy the system Eq. (\ref{Fourier}).

In the paper Ref. \onlinecite{WegnerHikami}, the
Hamiltonian Eq. (\ref{HSW}) was introduced to
describe the effect of a specific type of disorder
on electron states in $n=0$ LL. It was assumed that
the bare states were of $N=2$ types, and the
disorder scattering was allowed only between the states
of different type.
Upon examination
the expansion of the diffusion
coefficient in powers of disorder,
it was concluded that the state, $E=0$,
with $E$ measured from $\frac{eB\hbar}{2mc}$,
is delocalized. With regard to the density
of states, the self-consistent Born approximation\cite{Ando1974}
for the Hamiltonian Eq. (\ref{HSW}) yields a semicircle shape.
Taking  the large-$N$ limit, the authors concluded
that the density of states diverges logarithmically at $E=0$.
Later\cite{Lee1994},  upon employing the
semiclassical description, D. K. K. Lee demonstrated that, in addition to $E=0$
delocalized state, the model of Ref. \onlinecite{WegnerHikami} contains two additional
delocalized states of the conventional quantum Hall type.
Subsequent numerical simulations\cite{Hikami1996}
confirmed the existence of all three delocalized states, see however Ref. \onlinecite{ToInclude}.

In graphene, the role of states of two types, considered in Ref.~\onlinecite{WegnerHikami},
is played by the states at $K$ and $K'$ points, while the scattering between them
is provided by the bond disorder.

Below we propose an alternative approach to describe the eigenstates
of  Eq. (\ref{HSW}). We start by introducing the new variables
 \begin{equation}
\label{ab}
a_k=\frac{1}{2}\left(A_k+B_k\right),~~~b_k=\frac{1}{2}\left(A_k-B_k \right).
\end{equation}
With these variables, the system Eq. (\ref{System2}) takes the form
\begin{eqnarray}
Ea_{\kappa}-\sum_q (h_1)_{\kappa, q} a_q = -i\sum_q (h_2)_{\kappa,q} b_q,\label{System31} \\
Eb_{\kappa}+\sum_q (h_1)_{\kappa, q} b_q = i\sum_q (h_2)_{\kappa,q} a_q.\label{System32}
\end{eqnarray}
Our main assumption in analyzing the eigenstates of the system Eq. (\ref{System31})
is that the magnitudes and statistical  properties of $h_1(x,y)$ and $h_2(x,y)$ fields
are completely different. In particular, we assume that the magnitude of $h_2(x,y)$
is much smaller than the magnitude of $h_1(x,y)$ and treat $h_2(x,y)$ {\em perturbatively}.

\begin{figure}
\includegraphics[scale=0.5]{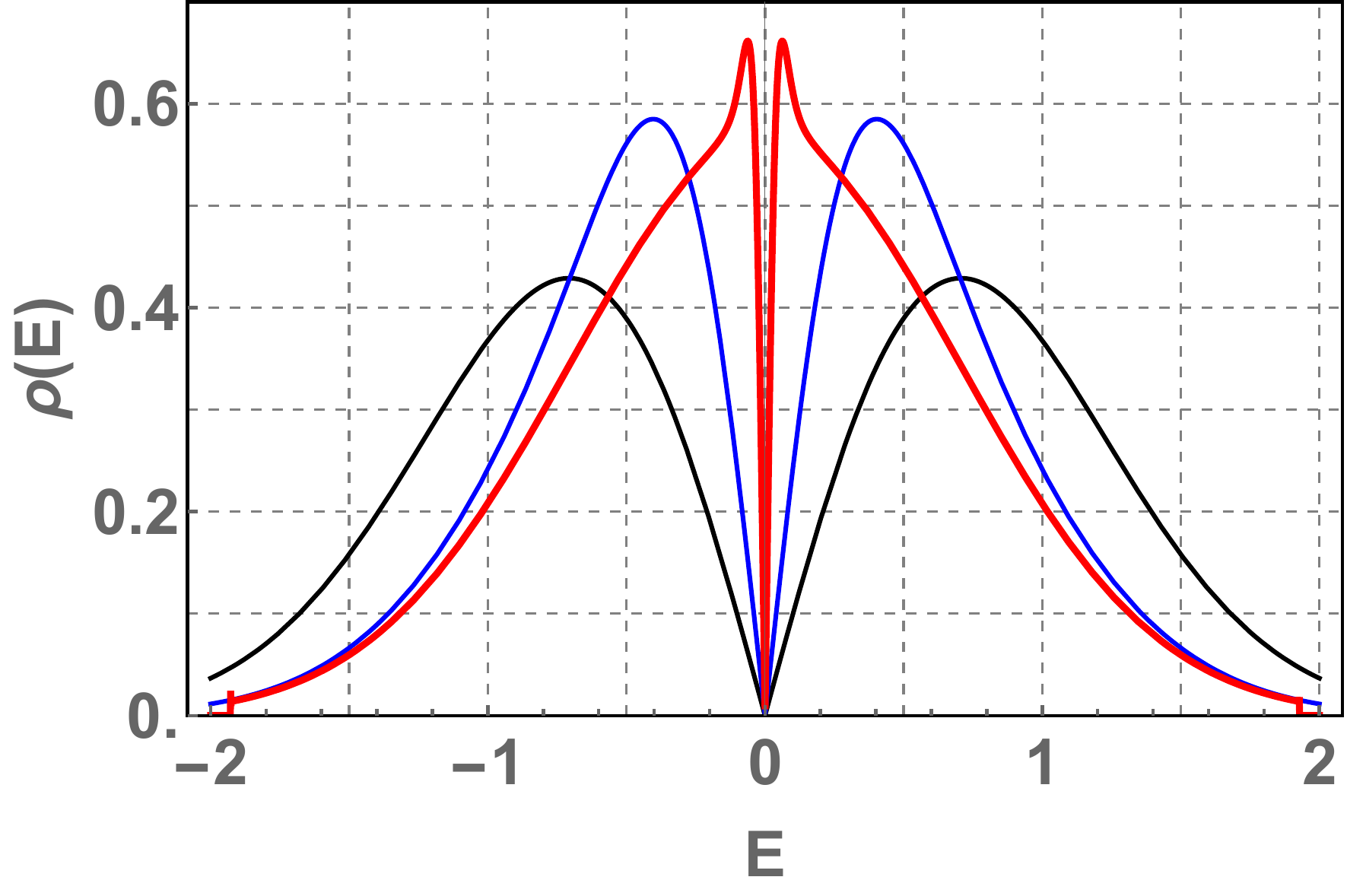}
\caption{(Color online) The shapes of the density of states are plotted from Eq. (\ref{FinalDOS}) versus the dimensionless energy, $E/\Gamma$, for three different values of the ratio $\gamma/\Gamma$: $1$ (black), $0.4$ (blue), $0.05$ (red). Black line corresponds to the Wigner-Dyson distribution. For small $\gamma/\Gamma$ a fine structure emerges near $E=0$.}
\label{fig2}
\end{figure}
In zeroth order, the eigenfunctions, $\chi_{\nu}^+(x,y)$ and $\chi_{\mu}^-(x,y)$,  of the system Eq. (\ref{System31})
are the states of $n=0$ LL in the {\em potentials} $h_1(x,y)$ and $-h_1(x,y)$, respectively.
Upon switching on the field $h_2(x,y)$,  the eigenfunctions, $\chi_{\nu}^+(x,y)$ and $\chi_{\mu}^-(x,y)$,
get coupled. The coupling amplitude is equal to
$\int d{\bm r} \left(\chi_{\nu}^+\right)^{\ast}h_2(x,y)\left(\chi_{\mu}^-\right)$. To proceed,
we further assume that the correlation length, $R_c$, of $h_2(x,y)$ is much bigger
than $\ell_{\s B}$. Then $h_2(x,y)$ in the integrand can be treated as a constant.
 Consequently, the coupling amplitude reduces to the overlap integral of $\chi_{\nu}^+$ and $\chi_{\mu}^-$.

Our prime observation is that this integral is nonzero only when $\chi_{\nu}^+$ and  $\chi_{\mu}^-$
correspond to energy $E$ in potential $h_1(x,y)$ and to $-E$ in potential $-h_1(x,y)$, respectively.
Then the functions  $\chi_{\nu}^+$ and  $\chi_{\mu}^-$ are the same. Any other function
$\chi_{\mu}^-$ in potential  $-h_1(x,y)$ has its counterpart in potential $h_1(x,y)$, which corresponds to energy different from the state $\chi_{\nu}^+$. Thus, it is orthogonal to $\chi_{\nu}^+$. This situation is illustrated in Fig. \ref{fig1}.

We conclude that, upon switching on the random field $h_2(x,y)$,
the modified states are determined upon diagonalizing the $2\times 2$ matrix
\begin{equation}
\label{2by2}
\begin{pmatrix}
E_{\nu} && i\left(h_2\right)_{\nu,\nu} \\ -i\left(h_2\right)_{\nu,\nu} && -E_{\nu}
\end{pmatrix},
\end{equation}
where $E_{\nu}$ is the bare energy, and the non-diagonal element, $\left(h_2\right)_{\nu,\nu}$, stands for the coupling amplitude.
The modified energies are given by
\begin{equation}
\label{modifiedE}
{\tilde E}_{\nu}=\pm \Big[E_{\nu}^2+|(h_2)_{\nu,\nu}|^2 \Big]^{1/2},
\end{equation}
Overall, the effect of $h_2(x,y)$ amounts to the repulsion of the states $E_{\nu}$ away from the center $E=0$.
\begin{figure}
\includegraphics[scale=0.4]{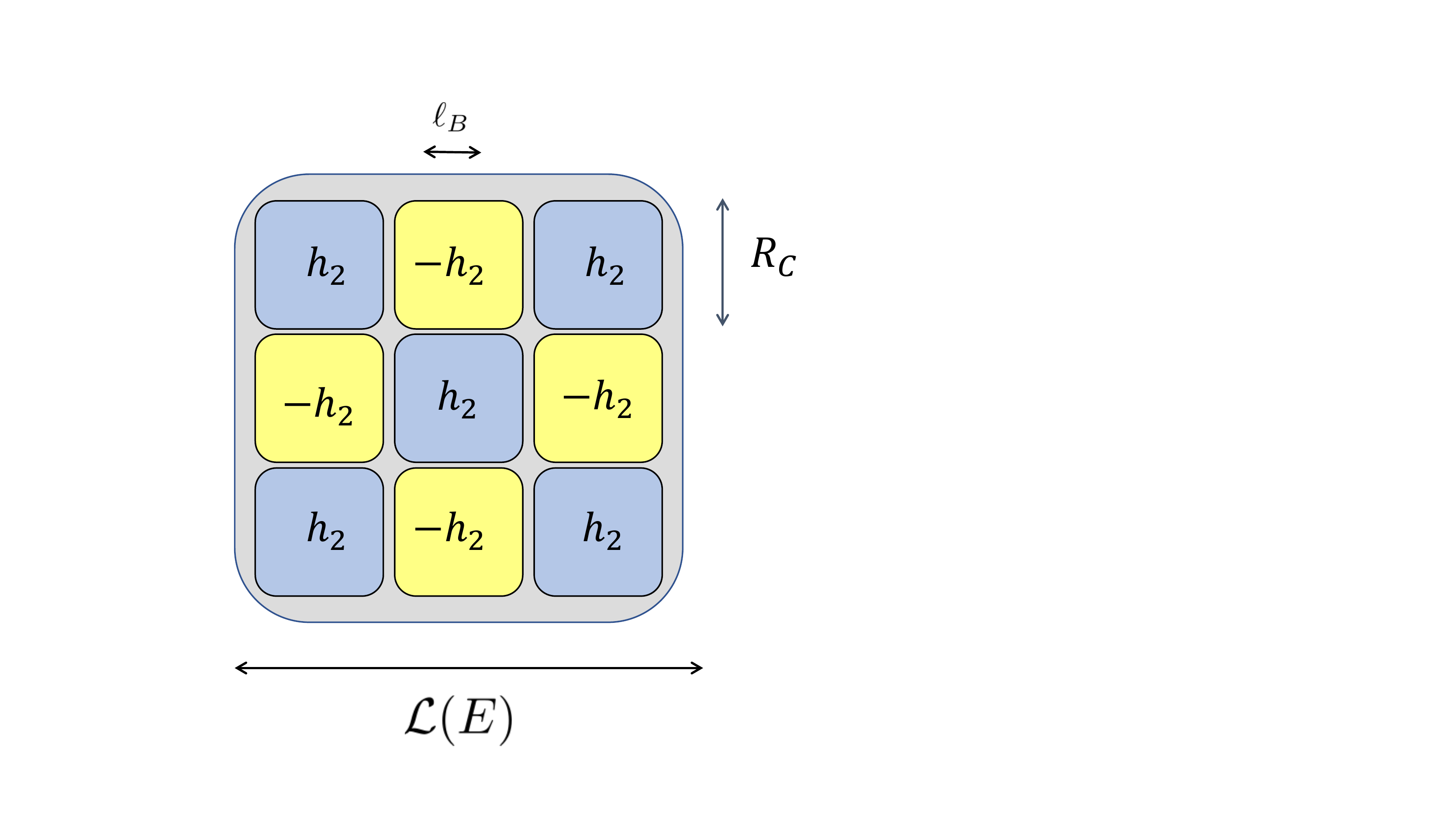}
\caption{(Color online) A cartoon illustrating the suppression of the
repulsion of energies $E$ and $-E$ near zero energy. Although the correlation
radius, $R_c$, is bigger than the magnetic length, $\ell_{\s B}$, the extension,
${\cal L}(E)$, of the wave-functions grows with decreasing $E$, and, eventually,
exceeds $R_c$. Then the matrix element, $(h_2)_{\nu,\nu}$, responsible for
repulsion, can be viewed as a sum of $\left(R_c/{\cal L}(E)\right)^2$ random contributions.}
\label{fig3}
\end{figure}
From the fact that the values $E_{\nu}$ and $(h_2)_{\nu,\nu}$ are
statistically independent, we readily arrive to the general expression
for the modified density of states
\begin{equation}
\label{Generalized}
\rho({\tilde E})=\int dE \rho_{\s h_1}(E)\int dh_2 {\mathlarger{\cal P}}(h_2) ~ \delta \Big[{\tilde E}-\left(E^2+ h_2^2\right)^{1/2} \Big],
\end{equation}
where $\rho_{\s h_1}(E)$ is the average density of states in the potential $h_1(x,y)$.
The form of this density of states depends on whether the correlation length of $h_1(x,y)$
is bigger or smaller than the magnetic length. For long-range disorder $\rho_{\s h_1}(E)$
is Gaussian
\begin{equation}
\label{rho1}
\rho_{\s h_1}(E)=\frac{1}{2\pi^{3/2}\ell_{\s B}^2\Gamma}\exp\left(-\frac{E^2}{\Gamma^2}\right),
\end{equation}
where the width, $\Gamma$, is simply $\Gamma_{\s L}=\langle h_1(x,y)^2\rangle^{1/2}$, i.e. the r.m.s. value of the potential.
In the opposite limit of short-range disorder
with a correlator
\begin{equation}
\label{correlator}
\langle h_1({\bm r})h_1({\bm r_1})\rangle = w \delta \left({\bm r}-{\bm r}_1\right)
\end{equation}
the exact density of states found by F. Wegner in Ref.~ \onlinecite{Wegner1983} has the form
\begin{figure}
\includegraphics[scale=0.4]{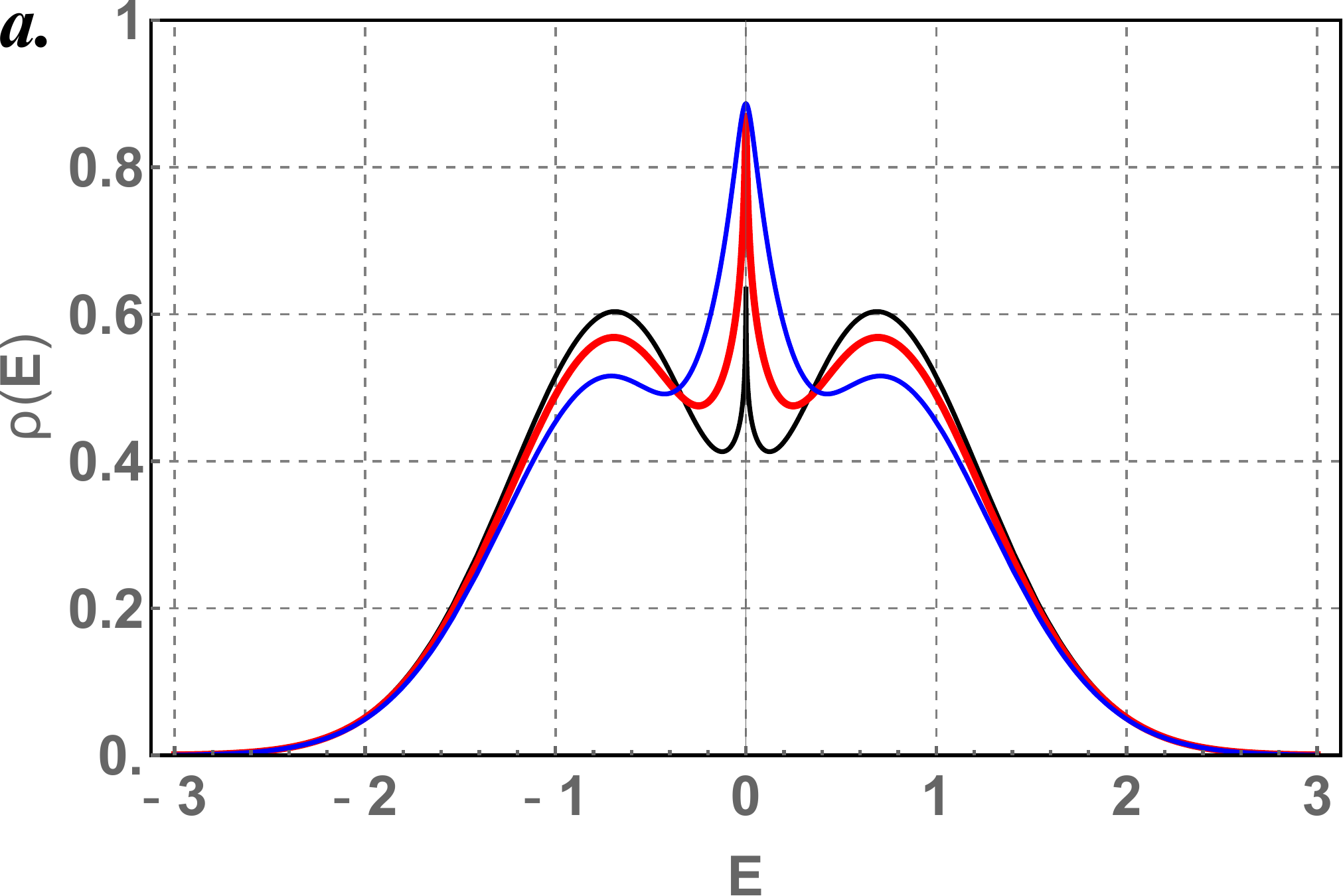}
\includegraphics[scale=0.4]{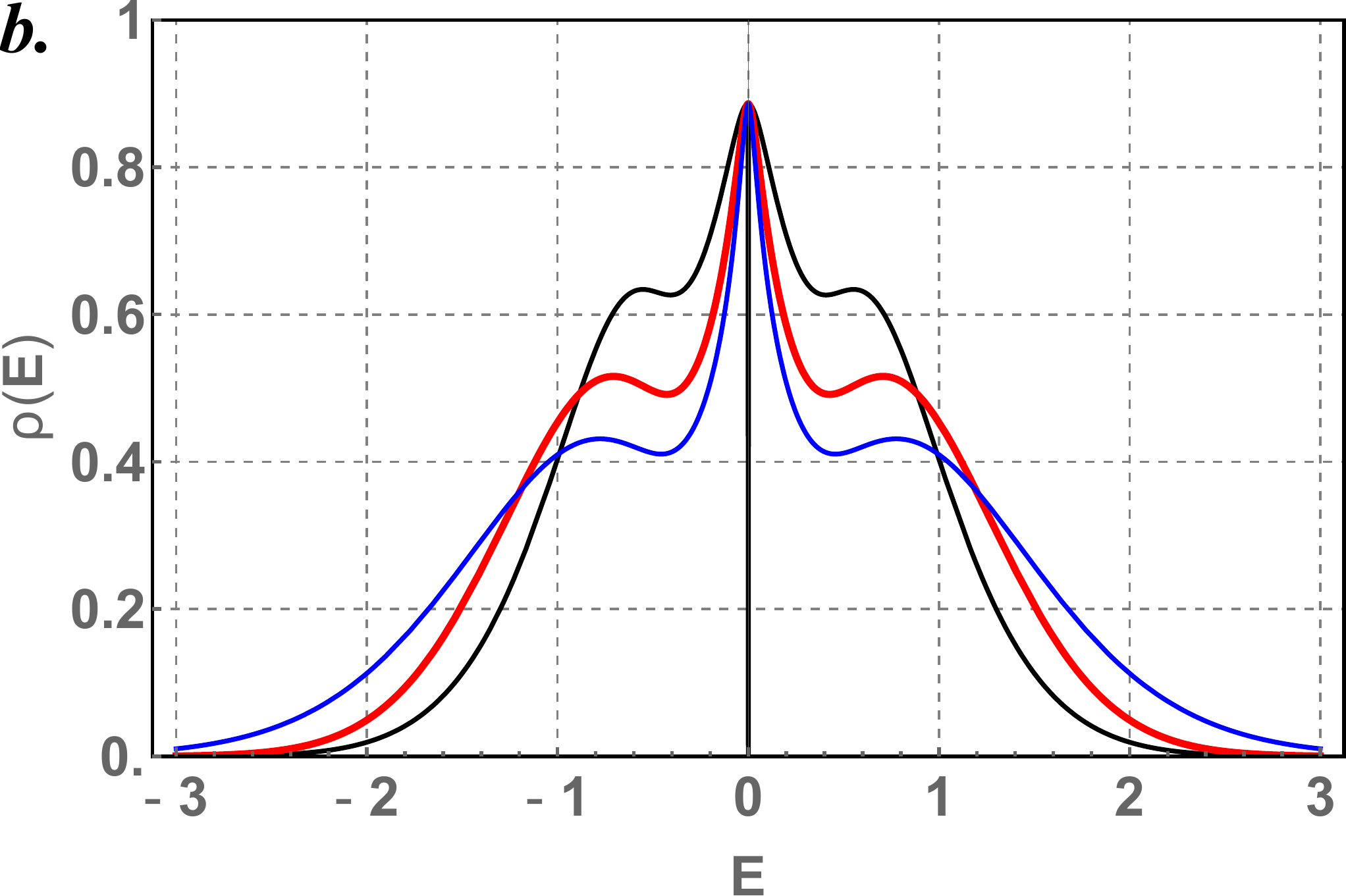}
\includegraphics[scale=0.4]{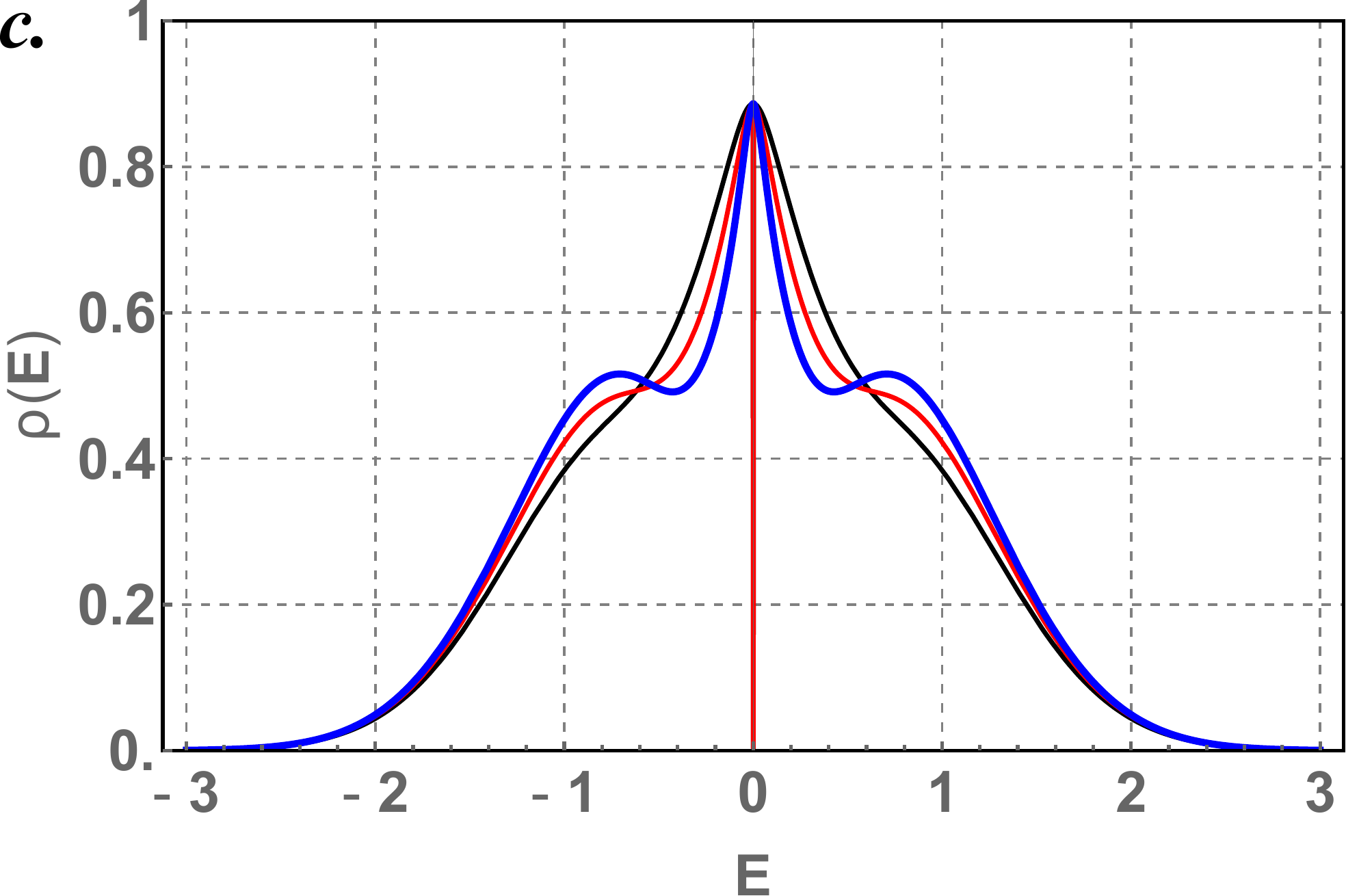}
\caption{(Color online) The density of states is plotted from Eqs. (\ref{Generalized3}), (\ref{interpolation})
 versus the dimensionless energy, $E/\Gamma$, by varying three parameters $\kappa$, $\gamma_0/\Gamma$, $R_c/\ell_{\s B}$. In the top panel $\rho(E)$ is plotted for three values of  $\kappa$: 1.2, (black) 1.5, (red) 2,~(blue) with $\gamma_0/\Gamma=1$, and $R_c/\ell_{\s B}=8$. In the middle panel (b), $\gamma_0/\Gamma$ takes values $0.5$, (black) $1$, (red) $1.5$, (blue) while $\kappa=2$ and $R_c/\ell_{\s B}=8$. In the bottom panel (c) $\rho(E)$ is  for the parameters $R_c/\ell_{\s B}$: $3$, (black) $5$, (red) $8$, (blue) with  $\gamma_0/\Gamma$ and $\kappa$ taking values $1$ and $2$, respectively.}
\label{fig4}
\end{figure}
\begin{equation}
\label{Wegner}
\rho_{\s h_1}(E)=\frac{1}
{2\pi^2\ell_{\s B}^2}\frac{\partial}{\partial E}\Big\{\arctan \Big[G\left(E/\Gamma \right)\Big]\Big\},~G(z)=\hspace{-1mm}\int_0^z\hspace{-1mm}dt~e^{t^2},
\end{equation}
with $\Gamma=\Gamma_{\s S} =\left(w/2\pi \ell_{\s B}^2\right)^{1/2}$.
While $\Gamma_{\s S}$ grows with magnetic field and $\Gamma_{\s L}$ does not,
the shape Eq. (\ref{Wegner}) is close to Gaussian.

Following our assumption that $h_2(x,y)$ is smooth, we choose the Gaussian form
for ${\mathlarger{\cal P}}(h_2)$
\begin{equation}
\label{Ph2}
{\mathlarger{\cal P}}(h_2)=\frac{1}{\pi^{1/2}\gamma}\exp\left(-\frac{h_2^2}{\gamma^2}\right),
\end{equation}
with $\gamma=\langle h_2(x,y)^2\rangle^{1/2}$.

Upon substituting Eq. (\ref{rho1}) and Eq. (\ref{Ph2}) into the expression
Eq. (\ref{Generalized}) for the density of states, we introduce polar coordinates $E=R \cos \varphi$ and
$h_2=R\sin\varphi$ and cast it in the form
\begin{multline}
\label{Generalized1}
\rho({\tilde E})=\frac{1}{2\pi^2\ell_B^2 \Gamma \gamma}\int\limits_{0}^{\infty}  dR~R  \int\limits_{0}^{2\pi} d\varphi \\
\times
 \exp\left[-R^2 \left(\frac{\cos^2 \varphi}{\Gamma^2}+\frac{\sin^2 \varphi}{\gamma^2}\right) \right]\delta({\tilde E}-R).
\end{multline}
Integration over $R$ is performed using the $\delta$-function, and the integral over $\varphi$ reduces to modified Bessel function, $I_0(z)$.
Final result reads

\begin{multline}
\label{FinalDOS}
\rho(\tilde{E}) =
 \frac{|\tilde{E}|}{\pi\ell_B^2\Gamma\gamma}
\exp\left[-\frac{\tilde{E}^2}{2}\left(\frac{1}{\Gamma^2}+\frac{1}{\gamma^2}\right)\right]
\\ \times I_0\left[\frac{\tilde{E}^2}{2}\left(\frac{1}{\Gamma^2}-\frac{1}{\gamma^2}\right)\right].
\end{multline}

The fact that $\rho({\tilde E})$ behaves as $|{\tilde E}|$ at small energies
is a natural consequence of the level repulsion. It is also natural that,
for symmetric disorder $\Gamma =\gamma$, Eq. (\ref{FinalDOS}) reduces to the
Wigner-Dyson distribution, as in Ref. \onlinecite{Hikami1996}. Under the assumption
adopted above, that the disorder $h_2(x,y)$ is weak, the shape of the density of
states develops a sharp feature at small energies, as illustrated in Fig. \ref{fig2},
which is somewhat reminiscent of the numerical data\cite{Chakravarty2007,Ando2007,Markos,Sweitzer,Japanese,Pereira,Lewenkopf,Roche2016},
but does not capture the robust low-energy behavior revealed in these papers.
We argue that the reason of the discrepancy lies in the fact that we disregarded the
energy dependence of the matrix element $(h_2)_{\nu,\nu}$.
Namely, when we assumed the correlation length, $R_c$, of $h_2(x,y)$ is much bigger than $\ell_{\s B}$, we overlooked the fact that, upon approaching to ${\tilde E}=0$
the eigenfunctions $\chi_{\nu}^+$ and $\chi_{\nu}^-$ become progressively extended.
\begin{multline}
\label{energydependent}
\langle\left( h_2\right)_{\nu,\nu}^2 \rangle=\int dr_1 |\chi_{\nu}^+(r_1)|^2dr_1
\int dr_2 |\chi_{\nu}^-(r_2)|^2dr_2
\\ \times
 \langle h_2(r_1) h_2(r_2) \rangle \sim \gamma \left(\frac{R_c}{{\cal L}(E)}\right)^2,
\end{multline}
where ${\cal L}(E)$ is the energy-dependent localization length of the wavefunctions  $\chi_{\nu}^+$ and $\chi_{\nu}^-$. We see that the repulsion of energy levels
from the band-center, $E=0$, gets strongly suppressed at $E\rightarrow 0$.
This observation is illustrated in Fig. \ref{fig3}.
The area corresponding to the ${\cal L}(E)^2$  state, $E$, contains $\left(\frac{{\cal L}(E)}{R_c}\right)^2$ squares within which $h_2({\bm r})$ is constant. Since the contributions of these squares to $(h_2)_{\nu,\nu}$
are random, the typical value of $(h_2)_{\nu,\nu}$ is suppressed by a factor
$\sim \left(\frac{R_c}{{\cal L}(E)}\right)$. This is certainly a hand-waving argument.
Strictly speaking, with $h_2({\bm r})$ changing in space, the state $\chi_{\nu}^-$ gets coupled to {\em all} the states $\chi_{\nu}^+$. It is, however, important that the
contributions to the matrix element from positive and negative energies almost
cancel each other at small $E$.
\subsection{Shapes of the density of states}

A minimal ansatz to incorporate the suppression of the repulsion
of the levels $E$ and $-E$ into the density of states Eq. (\ref{Generalized})
is to assume that the matrix element $(h_2)_{\nu,\nu}$ still obeys the
Gaussian distribution, but the
r.m.s. value, $\gamma$, is a function of $E$.

Due to finite $R_c$, the state, $\nu$, corresponding to the energy, $E$, will be coupled
by $h_2(x,y)$ not only to the state $-E$ but to the states corresponding to different energies.
We will still assume that only $(h_2)_{\nu,\nu}$ is non-zero, since the degree of violation
of the orthogonality is $\sim \ell_{\s B}^2/R_c^2$, which is small. Then, performing integration over $h_2$ in Eq. (\ref{Generalized}), we
arrive to the following expression for the density of states
\begin{multline}
\label{Generalized3}
\rho({\tilde E})=\frac{|{\tilde E}|}{2\pi\ell_{\s B}^2\Gamma}\int\limits_{0}^{{\tilde E}} \frac{dE}{\gamma(E)\left({\tilde E}^2-E^2 \right)^{1/2}}
\\
\times\exp\Bigg[-\Bigg(\frac{E^2}{\Gamma^2}+\frac{{\tilde E}^2-E^2}{\gamma^2(E)} \Bigg) \Bigg].
\end{multline}

Concerning the functional dependence of $\gamma(E)$, we know that it is constant
far from $E=0$ and falls off as $R_c/{\cal L}(E)$ as $E\rightarrow 0$.
To analyze the density of states, we chose the following interpolation:
\begin{equation}
\label{interpolation}
\gamma(E)=\gamma_0\tanh\left[\frac{R_c}{{\cal L}(E)}\right]=\gamma_0\tanh\left[\frac{R_c}{\ell_{\s B}}\left(\frac{E}{\Gamma}\right)^\kappa\right].
\end{equation}
Other forms of $\gamma(E)$ yielded similar results.
In fact, Eq. (\ref{Generalized3}) contains three independent parameters,
which we varied. The first is the strength, $\gamma_0$, of the disorder, $h_2(x,y)$,
as in Eq. (\ref{Ph2}), the second is the ratio $R_c/{\ell_{\s B}}$, which we assume
to be big, and, finally, the exponent, $\kappa$, in the energy dependence of the
localization length. For conventional quantum Hall critical point the value of $\kappa$
is $2.3$. In analysis of the shape, $\rho(E)$, we have changed one parameter keeping
the other two constant. The results are shown in Fig. \ref{fig4}. The main message of
Fig. \ref{fig4} is that, as we vary the parameters, the general shape of $\rho(E)$ remains unchanged.

From Fig. \ref{fig4}a we conclude that when the exponent $\kappa$ increases, the anomaly at $E=0$
becomes more and more pronounced. Comparing to Fig. \ref{fig2}, we see that the behavior of $\rho(E)\propto |E|$ gets modified to a narrow peak. The explanation for this is straightforward:
delocalization of states near $E=0$ in the absence of $h_2(x,y)$ results in suppression of  their
repulsion when $h_2(x,y)$ is switched on. This suppression becomes more effective upon increasing $\kappa$. Then the origin of the peak is that, while the states with $E\sim \Gamma$ are shifted by $h_2$ either to
the left or to the right, depending on the sign of $E$, the low-energy states retain their positions.
 Obviously, the analysis of the perturbation expansion in terms of $h_1$ and $h_2$, of the density of states
 up to a finite order, cannot capture this effect.
This is because the finite-order expansion does not capture the delocalization of the wave functions.

Fig. \ref{fig4}b suggests that, the prime effect of increasing the strength of $h_2$ is
the general broadening of the density of states, while the behavior at small $E$ changes weakly.

Evolution of the curves in Fig. \ref{fig4}c
can be understood as follows. We assumed that $h_2$ couples the state \
$\chi_{\nu}^+$ only to the state $\chi_{\nu}^-$. The bigger is $R_c$,
the more accurate is this assumption. Then, the bigger is $R_c$, the more
pronounced is the separation of the density of states into the central
peak and two split maxima.

From  all the curves in Fig. \ref{fig4} the most reminiscent
of the numerical simulation results\cite{Chakravarty2007,Ando2007,Markos,Sweitzer,Japanese,Pereira,Lewenkopf,Roche2016} is the red curve in Fig. \ref{fig4}a.
\subsection{White-noise disorder}
In this subsection we lift the requirement that the correlation radius, $R_c$, of the field $h_2(x,y)$ is much bigger than magnetic length. It
was this requirement that ensured the repulsion of the levels $E_{\nu}$
and $-E_{\nu}$. When $h_2(x,y)$ is short-ranged, it couples the
$K-K'$ level $E_{\nu}$ to all $K+K'$ levels, $E_{\mu}$.
Still, we will see that coupling of $E_{\nu}$ to $-E_{\nu}$ remains
distinguished, since the corresponding states have the same wave functions.

For this purpose we search for the solution of the system Eqs. (\ref{System31}), (\ref{System32})in the form of expansion,
\begin{equation}
\Psi^{\s (K+K')}=\sum\limits_{\nu} c_{\nu} \chi_{\nu}^+, ~ \Psi^{\s (K-K')}=\sum\limits_{\mu} d_{\mu} \chi_{\mu}^-,
\end{equation}
where $\chi_{\nu}^+$ and $\chi_{\mu}^-$ are the eigenfunctions of the system  Eqs. (\ref{System31}), (\ref{System32}) in the absence of $h_2(x,y)$. For a finite $h_2$ we arrive to the following system for the coefficients $c_{\nu}$ and $d_{\mu}$
\begin{eqnarray}
\label{pert}
c_{\nu} (E_{\nu}^+-E)\chi_{\nu}^+ +\sum\limits_{\nu'\neq\nu} c_{\nu'} (E_{\nu'}^+-E)\chi_{\nu'}^+\nonumber\\= -i h_2(x,y)\Bigg[d_{\mu}\chi_{\mu}^- +\sum\limits_{\mu'\neq\mu} d_{\mu'} \chi_{\mu'}^- \Bigg],\nonumber\\
d_{\mu} (E_{\mu}^--E)\chi_{\mu}^+ +\sum\limits_{\mu'\neq\mu}d_{\mu'} (E_{\mu'}^--E)\chi_{\mu'}^-\nonumber\\
=i h_2(x,y)\Bigg[c_{\nu}\chi_{\nu}^+ +\sum\limits_{\nu'\neq\nu}c_{\nu '} \chi_{\nu '}^+ \Bigg].
\end{eqnarray}
For a given $\nu$, we treat $c_{\nu}$ and $d_{\nu}$ as zero-order terms, and express $c_{\mu}$ and $d_{\mu}$ with $\mu \neq \nu$ through them. Substituting $c_{\mu}$, $d_{\mu}$ back into the system, we get
\begin{eqnarray}
\Bigg[E_{\nu}^+-E- \frac{|(h_2)_{\nu,\nu}|^2}{E_{\nu}^--E}\Bigg]c_{\nu}=d_{\nu} {\mathlarger S_{\nu}}
,\label{pert31}\\
\Bigg[E_{\nu}^--E- \frac{|(h_2)_{\nu,\nu}|^2}{E_{\nu}^+-E}\Bigg]d_{\nu}=-c_{\nu}{\mathlarger S_{\nu}},\label{pert32}
\end{eqnarray}
where ${\mathlarger S_{\nu}}$ stands for the sum
\begin{equation}
\label{S}
{\mathlarger S_{\nu}}=\sum\limits_{\mu\neq\nu} \frac{|(h_2)_{\nu,\mu}|^2}{E_{\mu}^--E}=\sum\limits_{\nu\neq\mu} \frac{|(h_2)_{\nu,\mu}|^2}{E_{\mu}^+-E}.
\end{equation}
Multiplying  Eq. (\ref{pert31}) and  Eq. (\ref{pert32}), we get the following equation for $E$
\begin{equation}
\label{quadratic}
\Big[\left(E^2-E_\nu^2 \right)-|(h_2)_{\nu,\nu}|^2\Big]^2=\left(E^2-E_\nu^2 \right) {\mathlarger S_{\nu}}^2,
\end{equation}
the solution of which reads
\begin{equation}
\label{solution}
E^2=E_\nu^2 +\Bigg[\left(|(h_2)_{\nu,\nu}|^2+ \frac{{\mathlarger S_{\nu}}^2}{4}\right)^{1/2}\pm\frac{{\mathlarger S_{\nu}}}{2} \Bigg]^2.
\end{equation}
This equation is a generalization of Eq. (\ref{modifiedE}).

 We can now estimate the accuracy of keeping only the diagonal elements of $h_2(x,y)$.
If the energy $E_{\nu}$ is in the ``body" of the band broadened by the potential $h_1(x,y)$, then only the neighboring  states contribute to $S$. This is because the
overlap with states at distance $x\gg \ell_{\s B}$ is small as $\exp(-x^2/2\ell_{\s B}^2)$. For a neighboring state, the typical value of the denominator is Eq. (\ref{S}) is $\sim \Gamma$, while the numerator is $\sim \gamma_0^2$. Thus, the relative correction to $(h_2)_{\nu,\nu}^2$ is $\sim \gamma_0^2/\Gamma^2$, i.e. it is small.

The estimate for the correction  ${\mathlarger S_{\nu}}$ in the case where $E_{\nu}\ll\Gamma$ should be carried out differently. With $h_2(x,y)$ being the white-noise, the average $\langle{\mathlarger S_{\nu}} \rangle$ contains the combination \begin{equation}
\label{combination}
{\mathlarger \sum \limits_{\mu}}\frac{|\chi_{\nu}^+(\bm{r})|^2 |\chi_{\mu}^-(\bm{r})|^2}{E_{\nu}^+ -E_{\mu}^-},
\end{equation}
which depends on the correlation between functions $\chi_{\nu}^+$ and $\chi_{\mu}^-$, which are the eigenfunctions in {\em different} potentials, $h_1(x,y)$ and $-h_1(x,y)$.

It is known, see e.g. Refs. \onlinecite{Chalkerscaling,Kravtsov,KravtsovPRL}, that the correlation of the critical eigenfunctions in the {\em same} potential is quantified as
\begin{equation}
\label{expression}
\int d\bm{r}~ |\chi_{\nu}^+(\bm{r})|^2|\chi_{\nu '}^+(\bm{r})|^2 \delta(E_{\nu}^+-E_{\nu '}^+) \propto|\frac{\Gamma}{E_{\nu}^+-E_{\nu '}^+}|^{\eta/2},
\end{equation}
where $\eta$ is the exponent characterizing the fractal structure of critical
eigenfunctions. Recall now, the wavefunction $\chi_{\nu}^+$ and $\chi_{\mu}^-$ are the same when they corresponds to opposite energies, $E_{\nu}^+=-E_{\mu}^-$. Then we can use Eq. (\ref{expression}) to estimate $\langle{\mathlarger S_{\nu}} \rangle$

\begin{equation}
\label{expression1}
\int \frac{dE_{\mu}^- ~ \rho_{\s h_1}(E_{\mu}^-)}{E_{\nu}^+-E_{\mu}^-}\Bigg|\frac{\Gamma}{E_{\nu}^++E_{\mu}^-} \Bigg|^{\eta/2}\propto\frac{1}{|E_{\nu}^+|^{\eta/2}}.
\end{equation}

The above equation suggests that $\langle{\mathlarger S_{\nu}} \rangle$ increases upon  approaching $E_{\nu}^+\rightarrow 0$. Still, it loses to the diagonal term, $(h_2)_{\nu,\nu}$. This is because a typical $(h_2)_{\nu,\nu}$ is proportional to $1/{\cal L}(E)$, see Eq. (\ref{energydependent}), while ${\mathlarger S_{\nu}}^2$ is proportional to $1/{\cal L}^2(E)$.
\section{Delocalized states}
The only physically transparent description of the quantum Hall transition
is the Chalker-Coddington (CC) network model of Ref. \onlinecite{CC}, which is a quantum generalization of the classical percolation. To apply this
model in our case, one should assume that both fields $h_1(x,y)$ and $h_2(x,y)$ are smooth. Then the semiclassical energies\cite{Lee1994} are determined by local values of $h_1$, $h_2$ and are equal to ${\cal E}_{\pm}=\pm \left(h_1^2+h_2^2\right)^{1/2}$. Within the prefactor, the distribution function, ${\cal F}({\cal E}_+)$, of ${\cal E}_{+}$  is given by Eq. (\ref{FinalDOS}). Then the percolation
threshold, $E=E_c$, is found from the condition $\int_0^{E_c}d{\cal E}_+\hspace{0.5mm}
{\cal F}({\cal E}_+) =\frac{1}{2}$. If $h_1$ and $h_2$ are statistically equivalent, then ${\cal F}({\cal E}_+)=\frac{2{\cal E}_+}{\Gamma^2}  \exp\left(-\frac{{\cal E}_+^2}{\Gamma^2} \right)$.  This yields $E_c=0.7\Gamma$, i.e. the delocalized state
lies slightly above the maximum, $0.5\Gamma$,  of the density of states. This is consistent with numerical result of Ref. \onlinecite{Hikami1996}, although the simulations were performed for the short-range disorder.

Classical percolation at $E=\pm E_c$ transforms into the conventional quantum Hall
transitions when the tunneling through the saddle points, defined by the conditions:
$\frac{\partial}{\partial x} (h_1^2+h_2^2)=0$ and
$\frac{\partial }{\partial y}(h_1^2+h_2^2)=0$
(and opposite signs of the second derivatives) are taken into account.

There is no classical picture underlying the delocalized state at $E=0$, revealed in Ref. \onlinecite{WegnerHikami}. A peculiar feature of this delocalization
established numerically in Refs. \onlinecite{Hikami1996}, \onlinecite{Markos}
is that the critical exponent is anomalously small, $\nu \approx 0.3$.
It is even smaller than the $\nu=\frac{4}{3}$ for classical percolation and for the random flux model.\cite{FurusakiImportant,ChalkerRandomFlux,ReadRandomFlux,BrouwerRandomFlux,ChiralCritPointSchweitzer}
It is likely that the accuracy of simulations on Refs. \onlinecite{Hikami1996}, \onlinecite{Chakravarty2007},
and \onlinecite{Markos}
was limited by the size effects.

\begin{figure}
\includegraphics[scale=0.3]{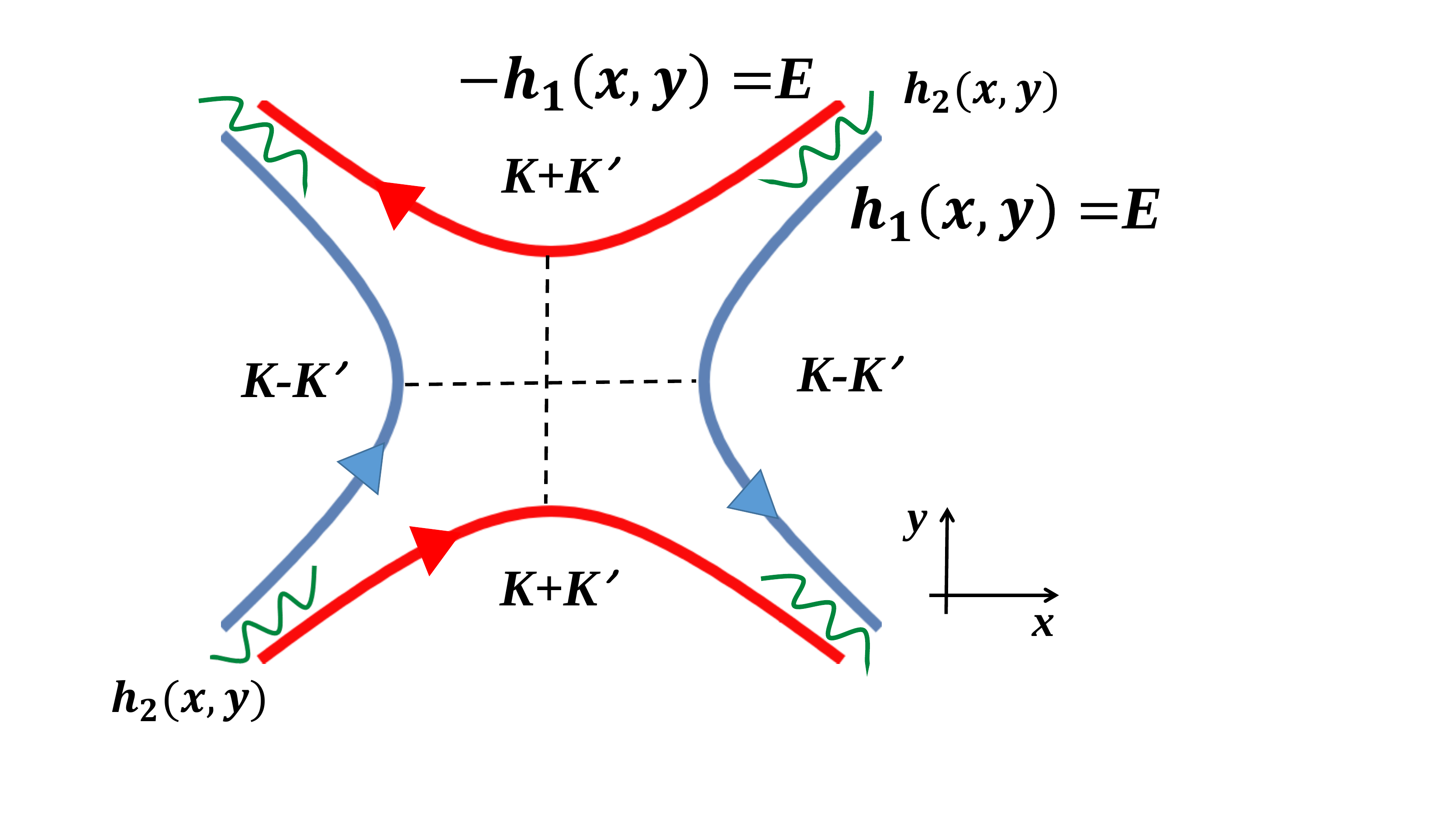}
\caption{(Color online) Blue equipotentials are $h_1(x,y)=E$, while red equipotentials
are $-h(x,y)=E$. Red and blue equipotentials are coupled via the random field $h_2(x,y)$.  Instead of tunneling through the saddle point, $\frac{\partial h_1}{\partial x} =0$, $\frac{\partial h_1}{\partial y} =0$, the transports proceeds as: Blue $\rightarrow$ Red $\rightarrow$  Blue and Red $\rightarrow$ Blue $\rightarrow$ Red. In this way, the saddle point gets bypassed.
The resulting energy dependence of the localization length originating from the   dependence of the transmission through the saddle point is weak.}
\label{fig5}
\end{figure}

Small critical exponent suggests that the localization length depends weakly on energy
near $E=0$. Below we invoke the CC model to explain a possible origin of
this weak dependence. The explanation is based on Fig.~\ref{fig5}. Within the CC model,
the behavior of the localization length on energy, $E$, is governed by tunneling via the saddle points separating two equipotentials, see Fig. \ref{fig5}. Equipotentials, $h_1(x,y)=0$, form a percolation network. Consider two blue equipotentials
corresponding to $K-K'$ states.
Note that, for the states corresponding to $K+K'$,
the random potential is equal to $-h_1(x,y)$. Thus, the equipotentials shown
in Fig. \ref{fig5} in red, are rotated by
$90^{\circ}$. Equipotentials $h_1(x,y)=E$ and $-h_1(x,y)=E$ are coupled by the random field, $h_2(x,y)$. This suggests that energy-dependent tunneling via the saddle point does not affect the structure of the low-energy states. The reason for this is that the saddle point is {\em bypassed}\cite{bypassing}
by the alternative channels:
blue $\rightarrow$ red $\rightarrow$  blue and red $\rightarrow$ blue $\rightarrow$ red, see Fig. \ref{fig4}.

\section{Discussion and concluding remarks}

({\em i}). By assuming that the bond disorder is strongly anisotropic, $h_2(x,y) \ll h_1(x,y)$,
we arrived to the following scenario for the shape of the density of states of the $n=0$
LL: the field $h_1(x,y)$ broadens the level into a band, while the field $h_2(x,y)$ is responsible for the repulsion of the levels {\em from the center of the band} facilitated by $K\rightarrow K'$ scattering. The states with
$E>0$ are shifted up, while the states with $E<0$ are shifted down. Most importantly,
the low-energy states remain unshifted, which leads to the three-peak structure of the
disorder-broadened band.

({\em ii}). Certainly, the assumption $h_2(x,y) \ll h_1(x,y)$ is artificial and does not
correspond to the simulations of Refs.
\onlinecite{Chakravarty2007,Ando2007,Markos,Sweitzer,Japanese,Pereira,Lewenkopf,Roche2016}.
However, treatment of $h_1(x,y)$ and $h_2(x,y)$ on the equal footing is possible
only within the self-consistent Born approximation, leading to the semicircle shape\cite{WegnerHikami}
with a width $\sqrt{2}\Gamma$. This means that the diagrams taken into account
within the self-consistent Born approximation\cite{Ando1974} do not capture properly the repulsion of the states away from the band center, $E=0$.
The picture of Ref.~\onlinecite{Lee1994} also does not
allow to make quantitative predictions about the shape of the
density of states near $E=0$.

The model of Ref. \onlinecite{WegnerHikami} is unique, in the sense, that delocalization of states
has a dramatic back effect on the density of states; self-consistent Born approximation is not sensitive to the localization. Also, evaluating any particular diagram in the perturbation expansion of the density of states
will not reveal an energy scale smaller than $\Gamma$. We inferred such a small scale from
delocalization of $K-K'$ and $K+K'$ eigenstates in the potential, $h_1(x,y)$, assuming that $h_2(x,y)$
is absent.
Note that, in the simulations of Refs. \onlinecite{Pereira}, \onlinecite{Lewenkopf}, the central peak in the density of states was hardly pronounced.
Accordingly, the authors did not find any evidence for delocalization at $E=0$.

The specifics of the Hamiltonian Eq. (\ref{HSW}) with regard to the behavior of the eigenstates at low energies was discussed in Ref. \onlinecite{Hikami1996}. It was pointed out that this specifics originates from the reflection symmetry of the Hamiltonian, $\sigma_z {\hat H}_{\s HSW} \sigma_z=-{\hat H}_{\s HSW}$.
Interestingly, the numerical simulations of a different model\cite{Zee}, 1D hopping chain with off-diagonal disorder,  described by a Hamiltonian possessing the reflection
symmetry, also revealed  a three-peak structure of the density of states.

({\em iii}) We have treated the field $h_2(x,y)$ perturbatively. This implies the  assumption
that the perturbation theory applies even at low energies, so that $h_2(x,y)$ does not modify the structure of the wave-functions of the low-energy states. On the other hand, the argument, illustrated in Fig. \ref{fig5},
suggests that the order in which $h_2$ and $E$ go to zero is important. This can be also seen from the analysis
of the expression Eq. (\ref{Generalized}) for the density of states. In the limit of low energies, the integral in Eq. (\ref{Generalized}) reduces to

\begin{equation}
\label{Asymptote}
\rho({\tilde{E}})\propto \int \limits_{0}^{\infty} \frac{dz}{z^{\kappa}} ~\delta\left(\frac{E}{E_c}-(z^2+z^{2\kappa})^{1/2} \right),
\end{equation}
where
\begin{equation}
E_c= \left(\frac{\Gamma^{\kappa}\ell_{\s B}}{\gamma_0 R_c}\right)^\frac{1}{1-\kappa}.
\end{equation}
Since $\gamma_0$ reflects the magnitude of $h_2$, it is seen that the result depends on the order
of taking the limits $h_2 \rightarrow 0$ and $E\rightarrow 0$.

({\em iv}) Naturally, in the opposite limit, $h_1(x,y) \ll h_2(x,y)$, we will arrive to the same result for the density of states and delocalization. In this limit, one should
 introduce the variables $A_k \pm iB_k$, instead of the variables $A_k\pm B_k$ Eq.(\ref{ab}). Then $h_2(x,y)$ will be responsible for broadening of the level,
 while $h_1(x,y)$ will lead to the repulsion of the states away from $E=0$.

({\em v}) Let as relate the fields $h_1(x,y)$ and $h_2(x,y)$ to the bond disorder in graphene: From Eq. (\ref{h}) we have

\begin{multline}
\label{h1h2}
h_1(\bm{r})=\sum_{bonds~i} \hspace{-3mm}c_i u(\bm{r}-\bm{r_i})\\ -\frac{1}{2}\hspace{-1mm}\Bigg[\sum_{bonds~ j} \hspace{-2mm} c_j  u(\bm{r}-\bm{r_j})+ \sum_{bonds~ l} \hspace{-2mm} c_l  u(\bm{r}-\bm{r_l})\Bigg],\\
h_2(\bm{r})=\frac{\sqrt{3}}{2}\Bigg[\sum_{bonds~ j} \hspace{-2mm} c_j  u(\bm{r}-\bm{r_j})- \sum_{bonds~ l} \hspace{-2mm} c_l  u(\bm{r}-\bm{r_l})\Bigg].
\end{multline}
Our analysis rests on the assumptions that $h_1$ is much bigger than $h_2$. Microscopically this means that the concentration of the perturbed $i$-bonds is much bigger than the concentration of the perturbed $j$ and $l$ bonds. In principle this situation can be realized in numerical simulations.
\section{Acknowledgements}

\vspace{1mm}
We are grateful to T. V. Shahbazyan for illuminating comments.
This work was supported by the Department of Energy,
Office of Basic Energy Sciences, Grant No.  DE- FG02-
06ER46313.



\begin{thebibliography}{30}

\bibitem{Ando1974} T. Ando and Y. Uemura,
``Theory of Quantum Transport in a Two-Dimensional Electron System under Magnetic Fields. I. Characteristics of Level Broadening and Transport under Strong Fields,"
J. Phys. Soc. Jpn. {\bf 36}, 959 (1974).



\bibitem{Baskin1978} {\'E}. M. Baskin, L. N. Magarill, and M. V. {\'E}ntin,
``Two-dimensional electron-impurity system in a strong magnetic field," Zh. Eksp. Teor. Fiz. {\bf 75}, 723 (1978) [Sov. Phys. JETP {\bf 48}, 365 (1978)].

\bibitem{Wegner1983} F. Wegner,
``Exact density of states for lowest Landau level in white noise potential superfield representation for interacting systems," Z. Phys. B {\bf 51}, 279 (1983).


\bibitem{Brezin1984} E. Br{\'e}zin, D. Gross, and C. Itzykson, ``Density of states in the presence of a strong magnetic field and random impurities," Nucl. Phys. {\bf B235}, 24 (1984).

\bibitem{Larkin1981} L. B. Ioffe and A. I. Larkin, ``Fluctuation levels
and cyclotron resonance in a random potential,"  Zh. Eksp. Teor. Fiz. {\bf 81}, 1048 (1981)
[Sov. Phys. JETP {\bf 54}, 556 (1981)].

\bibitem{Affleck1983} I. Affleck,
``Two-dimensional disorder in the presence of a uniform magnetic field,"  J. Phys. C{\bf 16}
 5839 (1983).

\bibitem{Affleck1984} I. Affleck, ``Density of states in a uniform magnetic field
and a white noise potential," J. Phys. C{\bf 17}, 2323 (1984).

\bibitem{Apel1987} W. Apel, ``Asymptotic density of states for a disordered 2D
electron system in a strong magnetic field," J. Phys. C{\bf 20} L577 (1987).

\bibitem{Chalker1985}  K. A. Benedict and J. T. Chalker,
``Properties of the disordered two-dimensional electron gas in a strong magnetic field,"
J. Phys. C{\bf 18}, 3981 (1985).


\bibitem{Chalker1986} K. A. Benedict and J. T. Chalker,
``An exactly solvable model of the disordered two-dimensional electron gas in a strong magnetic field,"
J. Phys. C {\bf 19}, 3587 (1986).


\bibitem{Benedict1987} K. A. Benedict,``The fate of the Lifshitz tails of high Landau levels,"
Nucl. Phys. {\bf B280}, 549 (1987).


\bibitem{Shahbazyan1993} M. E. Raikh and T. V. Shahbazyan, ``High Landau levels in a smooth
random potential for two-dimensional electrons," Phys. Rev. B {\bf 47}, 1522 (1993).



 \bibitem{Ando2002} Y. Zheng and T. Ando, ``Hall conductivity of a two-dimensional graphite system,"
Phys. Rev. B. {\bf 65} 245420 (2002).


 \bibitem{ReviewOnLLsInGraphene} M. O. Goerbig, ``Electronic properties of graphene in a strong magnetic field," Rev. Mod. Phys. {\bf 83}, 1193 (2011).



\bibitem{StormerKim} Z. Jiang, Y. Zhang, H. L. Stormer, and P. Kim,
``Quantum Hall States near the Charge-Neutrality Dirac Point in Graphene,"
Phys. Rev. Lett.  {\bf 99}, 106802 (2007).


\bibitem{Geim} L. A. Ponomarenko, R. Yang, R.V. Gorbachev, P. Blake, A. S. Mayorov, K. S. Novoselov, M. I. Katsnelson, and A. K. Geim, ``Density of States and Zero Landau Level Probed through Capacitance of Graphene," Phys. Rev. Lett. {\bf 105}, 136801 (2010).


\bibitem{JapaneseExperimental} K. Takase, H. Hibino, and K. Muraki, ``Probing the extended-state width of disorder-broadened Landau levels in epitaxial graphene," Phys. Rev. B {\bf 92}, 125407 (2015).


\bibitem{Gornyi2008} P. M. Ostrovsky, I. V. Gornyi, and A. D. Mirlin, ``Theory of anomalous quantum Hall effects in graphene," Phys. Rev. B. {\bf 77}, 195430 (2008).

 \bibitem{DisorderLL} W. Zhu, Q.W. Shi, X. R. Wang, J. Chen, J. L. Yang, and J. G. Hou, ``Shape of Disorder-Broadened Landau Subbands in Graphene," Phys. Rev. Lett. {\bf 102}, 056803 (2009).








\bibitem{Chakravarty2007} P. Goswami, X. Jia, and
S. Chakravarty, ``Quantum Hall plateau transition in the lowest Landau level of disordered graphene," Phys. Rev. B {\bf 76}, 205408 (2007).


\bibitem{Ando2007} M. Koshino and T. Ando, ``Splitting of the quantum Hall transition in disordered graphenes," Phys. Rev. B {\bf 75}, 033412 (2007).



\bibitem{Markos}
P. Marko{\v s}, and L. Schweitzer
``Critical conductance of two-dimensional chiral systems with random magnetic
flux," Phys. Rev. B
{\bf 76}, 115318 (2007).





\bibitem{Sweitzer} L. Schweitzer, ``Narrow depression in the density of states at
the Dirac point in disordered graphene,"
Phys. Rev. B {\bf 80}, 245430 (2009).

\bibitem{Japanese} T. Kawarabayashi, Y. Hatsugai, and H. Aoki, ``Quantum Hall Plateau Transition in Graphene with Spatially Correlated Random Hopping," Phys. Rev. Lett. {\bf 103}, 156804 (2009).





\bibitem{Pereira} A. L. C. Pereira, ``Splitting of critical energies in the $n=0$
Landau level of graphene," J. Phys. C {\bf 11} 095019  (2009).

\bibitem{Lewenkopf} A. L. C. Pereira, C. H. Lewenkopf, and E. R. Mucciolo, ``Correlated random hopping disorder in graphene at high magnetic fields: Landau level broadening and localization properties," Phys. Rev. B {\bf 84}, 165406 (2011).



\bibitem{Roche2016} N. Leconte, F. Ortmann, A. Cresti, and S. Roche,
``Unconventional features in the quantum Hall regime of disordered graphene: Percolating impurity
states and Hall conductance quantization," Phys. Rev. B {\bf 93}, 115404 (2016).


\bibitem{Atiyah} Y. Aharonov and A. Casher, ``Ground state of a spin-$1/2$ charged particle in a two-dimensional magnetic field," Phys. Rev. A {\bf 19}, 2461 (1979).

\bibitem{WegnerHikami} S. Hikami, M. Shirai, and F. Wegner,
``Anderson localization in the lowest Landau level
for a two-subband model,"
Nucl. Phys. {\bf B408}, 413 (1993).


\bibitem{Lee1994} D. K. K. Lee, ``Degenerate Landau bands with interband disorder: A semiclassical picture." Phys. Rev. B {\bf 50}, 7743 (1994).

\bibitem{Hikami1996} K. Minakuchi and S. Hikami, ``Numerical study of localization in the two-state Landau level,"
Phys. Rev. B {\bf 53}, 10898 (1996).


\bibitem{ToInclude} C. B. Hanna, D. P. Arovas, K. Mullen, and S. M. Girvin,
``Effect of spin degeneracy on scaling in the quantum Hall regime,"
Phys. Rev. B {\bf 52}, 5221 (1995).

\bibitem{Chalkerscaling} J. T. Chalker and G. J. Daniell, ``Scaling, Diffusion, and the Integer Quantized Hall Effect," Phys. Rev. Lett. {\bf 61}, 593 (1988).

\bibitem{KravtsovPRL} M. V. Feigel'man, L. B. Ioffe, V. E. Kravtsov, and E. A. Yuzbashyan, ``Eigenfunction Fractality and Pseudogap State near the Superconductor-Insulator Transition," Phys. Rev. Lett. {\bf 98}, 027001 (2007).

\bibitem{Kravtsov} V. E. Kravtsov, A. Ossipov, O. M. Yevtushenko, and E. Cuevas, ``Dynamical scaling for critical states: Validity of Chalker's ansatz for strong fractality," Phys. Rev. B {\bf 82}, 161102(R) (2010).




\bibitem{CC} J. T. Chalker and P. D. Coddington, ``Percolation, quantum tunneling, and the integer quantum Hall effect," J. Phys. C {\bf 21}, 2665 (1988).







\bibitem{FurusakiImportant} A. Furusaki,
``Anderson Localization due to a Random Magnetic Field in Two Dimensions,"
Phys. Rev. Lett. {\bf 82}, 604 (1999).



\bibitem{ChalkerRandomFlux} V. Kagalovsky, B. Horovitz, Y. Avishai, and J. T. Chalker,
``Quantum Hall Plateau Transitions in Disordered Superconductors,"
Phys. Rev. Lett. {\bf 82}, 3516 (1999).




\bibitem{ReadRandomFlux}
I. A. Gruzberg, A. W. W. Ludwig, and N. Read,
``Exact Exponents for the Spin Quantum Hall Transition,"
Phys. Rev. Lett. {\bf 82}, 4524 (1999).




\bibitem{BrouwerRandomFlux} C. Mudry, P. W. Brouwer, and A. Furusaki, ``Random magnetic flux problem in a quantum wire,"
Phys. Rev. B {\bf 59} 13221 (1999).


\bibitem{ChiralCritPointSchweitzer} L. Schweitzer and P. Marko{\v s}, ``Scaling at chiral quantum
critical points in two dimensions,"
Phys. Rev. B {\bf 85}, 195424 (2012).


\bibitem{bypassing} D. G. Polyakov and M. E. Raikh, ``Quantum Hall Effect in Spin-Degenerate Landau Levels: Spin-Orbit Enhancement of the Conductivity,"
Phys. Rev. Lett. {\bf 75}, 1368  (1995).


\bibitem{Zee} S. Hikami, A. Zee, ``Complex random matrix models with possible
applications to spin-impurity scattering in quantum
Hall fluids," Nucl. Phys. {\bf B446}, 337 (1995).


%
%
%
%

























\end{thebibliography}
\end{document}